\documentclass[conference]{IEEEtran}

\usepackage{graphicx} % Required for inserting images
\usepackage{colortbl} 
\usepackage{xcolor}   % For color definitions
\usepackage{fullpage}
\usepackage{subcaption}
\usepackage{xcolor}
\usepackage{amsfonts,amsmath,amssymb,amsthm}
\usepackage{hyperref}
\usepackage{cite}
\usepackage{breqn}
\usepackage{mathtools}
\usepackage{babel}
\usepackage{xcolor}
\usepackage{soul}

\begin{document}

\title{User Authentication and Vital Signs Extraction from Low-Frame-Rate and Monochrome No-contact Fingerprint Captures}

\author{
\IEEEauthorblockN{Olaoluwayimika Olugbenle, Logan Drake, Naveenkumar G. Venkataswamy, Arfina Rahman, Yemi Afolayanka}
\IEEEauthorblockA{\textit{Dept. of Electrical and Computer Engineering} \\
\textit{Clarkson University}\\
Potsdam, NY \\
\{olugbeoj, drakelj, venkatng, arahma, afolayyj\}@clarkson.edu}
\and
\IEEEauthorblockN{Masudul Imtiaz}
\IEEEauthorblockA{\textit{Dept. of Electrical and Computer Engineering} \\
\textit{Clarkson University}\\
Potsdam, NY \\
mimtiaz@clarkson.edu}
\and
\IEEEauthorblockN{Mahesh K. Banavar}
\IEEEauthorblockA{\textit{Dept. of Electrical and Computer Engineering} \\
\textit{Clarkson University}\\
Potsdam, NY \\
mbanavar@clarkson.edu}
}

\maketitle

\begin{abstract}
We present our work on leveraging low-frame-rate monochrome (blue light) videos of fingertips, captured with an off-the-shelf fingerprint capture device, to extract vital signs and identify users. These videos utilize photoplethysmography (PPG), commonly used to measure vital signs like heart rate. While prior research predominantly utilizes high-frame-rate, multi-wavelength PPG sensors (e.g., infrared, red, or RGB), our preliminary findings demonstrate that both user identification and vital sign extraction are achievable with the low-frame-rate data we collected. Preliminary results are promising, with low error rates for both heart rate estimation and user authentication. These results indicate promise for effective biometric systems. We anticipate further optimization will enhance accuracy and advance healthcare and security.
\end{abstract}

\section{Introduction}
\label{sec:intro}

One of the main challenges to fingerprint-based authentication systems is the problem of biometric spoofing. One set of solutions is liveness detection, where the fingerprint sensor detects whether the finger placed on the sensor is live. In our approach, we demonstrate liveness by calculating vital signs such as heart rate, respiratory rate, and oxygen saturation using photoplethysmography (PPG). Since the PPG can be extracted with technology that can be co-located with fingerprint sensors, it is a natural choice for liveness detection, providing multiple signs of liveness. 

\begin{figure}[!t]
    \centering
    \includegraphics[width=0.4\textwidth]{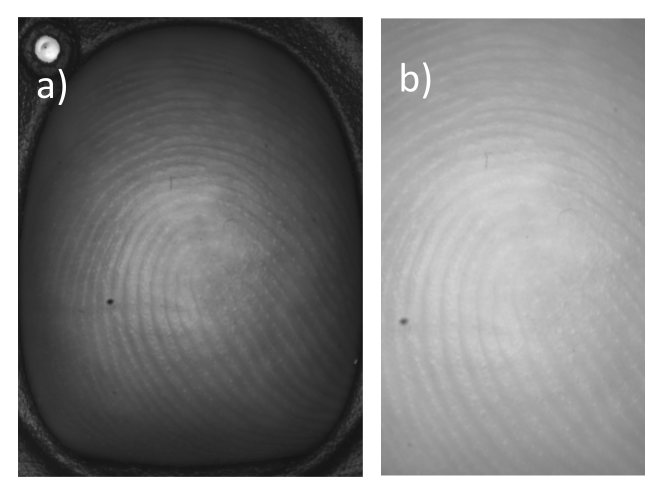}
    \caption{a) Originally captured image. b) Image after cropping. We use an off-the-shelf no-contact fingerprint sensor that collects blue-light 3000ppi images at a low frame rate of 14 frames per second. The image obtained from the sensor is cropped to retain only the image of the fingertip and eliminate background artifacts.}
    \label{fig:image_crop}
\end{figure}

\begin{figure*}[!tb]
    \centering
    \includegraphics[width=0.9\textwidth]{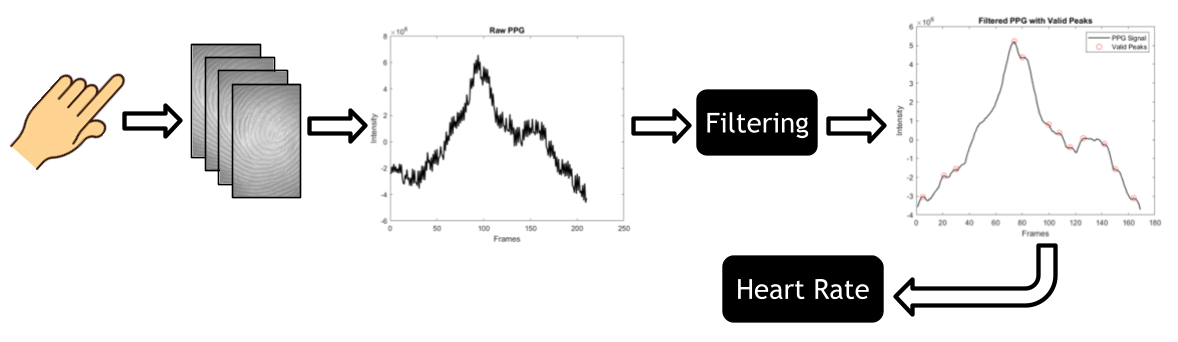}
    \caption{Extracting the heart rate from the PPG. The algorithm is described in Section \ref{sec:Vitals}.}
    \label{fig:PPG_Background}
\end{figure*} 

PPG is a non-invasive, light-based method for measuring the changes in blood volume at the microvascular bed of tissue. As the heart beats, the volume of blood at the body's extremities (e.g., fingertips) varies periodically. Since blood absorbs light, observing how the light intensity varies at the fingertip gives us an indirect measure of several vital signs including heart rate and oxygen saturation. 

Most medical PPGs use red light and infrared (IR) frequencies to detect heart rate and oxygen saturation. This can also be extended to RGB videos \cite{ayesha2021heart}, with the red channel used for heart rate estimation and the red and blue channels used for oxygen saturation.

Due to individual variations of the capillaries in the fingertip, it has been hypothesized that the PPG can be used for user authentication \cite{bastos2021smart, yalan2021ppg, Jindal2016adaptive, Lovisotto2020seeing}.  Methods include the use of statistical tools to generate a ``Human ID'' \cite{bastos2021smart} and deep learning pipelines with CNNs and LSTMs \cite{Everson2018biometricnet}.

Existing methods for vital signs extractions and user authentication use high frame-rate red/IR or RGB images/videos. However, in this paper, we use an off-the-shelf no-contact fingerprint sensor that uses blue light only and collects 3000ppi images at a low frame rate of 14 frames per second.  This poses challenges due to blue light and low frame rate. We adapt existing methods developed for red/IR and RGB to work with the data collected by the fingerprint sensor we use. 

The remainder of this paper is organized as follows: Section \ref{sec:Background} gives more details on the COTS camera and explores the process for extracting PPG signals. Section \ref{sec:Vitals} explores the methods used for extracting vital signs and the results. Section \ref{sec:ID} investigates how user identification was carried out and its results. Finally, Section \ref{sec:conclusions} summarizes the work done in this paper, and provides future directions for exploration. 

\section{Fingerprint sensor and PPG extraction}
\label{sec:Background}

We use an off-the-shelf no-contact fingerprint sensor that collects blue-light 3000ppi images at a low frame rate of 14 frames per second. 
The device is USB-enabled and is designed for the seamless collection of a continuous stream of fingerprint images. Its design minimizes collection variability by securely fixing the finger's position, ensuring proper focus and resolution, while also mitigating the impact of background variations. 

We obtain 15-20 seconds worth of images from a user. The user places their finger entirely over the camera's field-of-view (FOV) while the blue light source illuminates the finger. Once these images are obtained from the COTS device, we crop each one to retain only the fingertip image, as shown in Fig. \ref{fig:image_crop}. 
We calculate the total intensity of the pixels for each image (or frame) and plot this across all frames to yield the PPG signal. 
These signals are then processed to find the user's vital signs (e.g. heart rate) as shown in Fig. \ref{fig:PPG_Background}. 

\section{Vital Signs Extraction}
\label{sec:Vitals}

Our approach to liveness detection is to extract the vital signs of the user. To confirm that a person is alive, two criteria must be met: first, the individual must exhibit a measurable heart rate, and second, this heart rate must fall within the acceptable range for a living human being. In this work, we focus on extracting the user's vital signs from video recordings using the PPG extraction process described in Section \ref{sec:Background}.

The first step in this process is to clean and filter the PPG signal. We detrend the signal to remove any polynomial drift and DC shift. 

In the next step, we apply a 6th-order Chebyshev low-pass filter. From this filtered signal, we can calculate the heart rate of the user. Using the frame rate of the video (14 frames per second for the COTS device we used) and the number of frames between successive peaks, the time between two successive peaks can be calculated. Using this, the heart rate, measured in beats per minute, can be calculated as
\begin{equation}
\textrm{Heart Rate} = \frac{60}{\textrm{Median time between peaks}}. 
\label{eqn:HR_calc}
\end{equation}

\begin{figure*}[ht]
    \centering
    \includegraphics[width=0.82\textwidth]{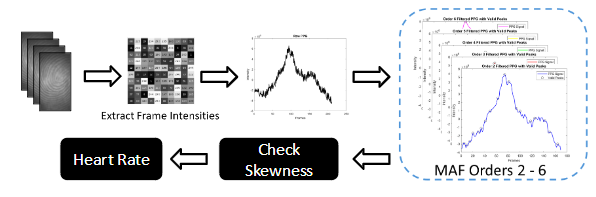}
    \caption{Grayscale PPG Pipeline.}
    \label{fig:Vital_Extraction_pipeline}
\end{figure*} 

\subsection{Improving Accuracy}
\label{ssec:PPG_Updated_Algorithm}

To enhance the accuracy of heart rate estimation, a more advanced filtering approach was developed. The raw PPG signal is run through 5 separate moving average filters, with their orders increasing from the 2nd to the 6th order. The heart rate is calculated from the output of each filter using (\ref{eqn:HR_calc}), resulting in 5 individual heart rate estimates. In addition to the median, the skewness is also calculated, 
which is used to check the quality of the obtained PPG signal. If the skewness is less than 13\%, the signal is deemed good, and the mean of the 5 calculated heart rates can be used as the most accurate estimation. If the skewness is greater than 13\% the signal is considered noisy and the heart rate calculated from the second-order moving average filter is used as the most accurately estimated heart rate. 

\subsection{Results}

For testing, we record PPG signals over 6 different trials for 15 seconds across 5 unique users, collecting 30 distinct PPG signals. 

The flow described in Section \ref{ssec:PPG_Updated_Algorithm}
shows improvement over just the method in (\ref{eqn:HR_calc}). 
% our preliminary results. 
Our preliminary algorithm shown in (\ref{eqn:HR_calc}) resulted in a mean percent error of 22.9\%. With the improvements discussed in section \ref{ssec:PPG_Updated_Algorithm}, the mean percent error reduces to 11.6\%. As an illustration, the results for user 3 are shown in Table \ref{tab:vital_signs_results}.

\begin{table}[!ht]
\centering
\caption{Ground truth heart rate (HR) values compared to our algorithm’s calculated values for ``User 3''. Improved heart rate estimates are obtained using the algorithm described in Section \ref{ssec:PPG_Updated_Algorithm}.}
    \begin{tabular}{|l|l|l|l|}
    \hline
    \rowcolor{black} % Header row color
\multicolumn{1}{c|}{\textcolor{white}{Trial}} & \multicolumn{1}{c|}{\textcolor{white}{\shortstack{Ground Truth \\ (Beats/Min)}}} & 
\multicolumn{1}{c|}{\textcolor{white}{\shortstack{Calculated \\ \textcolor{white}{(Beats/Min)}}}} &
\multicolumn{1}{c|}{\textcolor{white}{Error (\%)}}
\\ \hline
        Trial 1 & \centering 91 & 96.14 & 5.7\% \\ \hline        \rowcolor{lightgray} 
        Trial 2 & 87 & 105.66 & 21.4\% \\ \hline
        Trial 3 & 85 & 79.64 & 6.3\% \\ \hline        \rowcolor{lightgray} 
        Trial 4 & 84 & 79.85 & 4.9\% \\ \hline
        Trial 5 & 86 & 85.44 & 0.7\%  \\ \hline
        \rowcolor{lightgray} 
        Trial 6 & 89 & 80.47 & 9.6\% \\ \hline        
        \rowcolor{white} 
        ~ & ~ & Average \% Error  & 8.1\%\\ \hline
        \rowcolor{green}
        ~ & ~ & Average \% Error  & 11.6\%   \\ 
        \rowcolor{green}
        ~ & ~ & for all users & \\ \hline
    \end{tabular}
\label{tab:vital_signs_results}
\end{table}

\section{User Identification}
\label{sec:ID}

Since the blood flow to the fingertip depends on the cardiovascular system and capillaries, the PPG signals derived as a result of blood flowing through these systems are unique to each user and can be relied on for user authentication \cite{Lovisotto2020seeing}. While using PPG signals for authentication has been demonstrated with red/IR and RGB-derived PPG signals \cite{bastos2021smart, Everson2018biometricnet}, here, we show our first attempts at using blue monochrome low frame-rate PPG signals for authentication. For preliminary results, we (1) adapt the Human ID system from \cite{bastos2021smart}, and (2) use a system comprised of deep networks, including convolutional neural networks (CNNs) and long short-term memory networks (LSTMs).

\begin{figure*}[ht]
    \centering
    \includegraphics[width=0.8\textwidth]{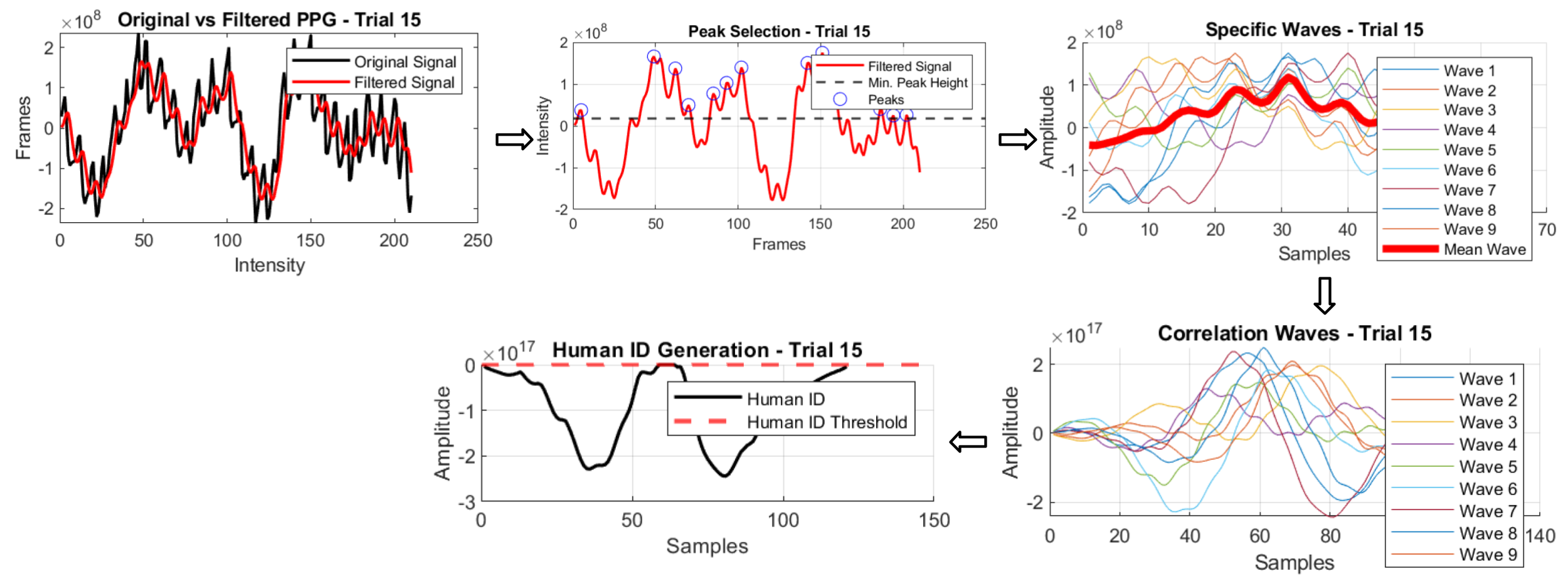}
    \caption{Human ID Pipeline.}
    \label{fig:human_ID_pipeline}
\end{figure*} 

\subsection{Human ID}
\label{ssec:Human_ID}

In this method, we adapt the statistical tools utilized in \cite{bastos2021smart} to develop a human ID that uniquely represents a user based on their PPG signal.

\subsubsection{Algorithm}
\label{sssec:human_id_algorithms}

Our preliminary work is based on the pipeline introduced by the authors in \cite{bastos2021smart}. Initially, the PPG data is filtered using a 2nd-order Chebyshev low-pass filter to de-noise the signal without smoothing out peaks in the signal. From the filtered data, we select peaks, which will then be used for the identification process. 
After peak selection, we extract peaks, taking a few samples to the left and right of the peak, forming a wave. We drop any peaks near the start or end of the signal. This process of wave formation is then repeated for all the valid peaks found. Afterward, the mean of all the waves is found. Next, the cross-correlation between each wave and the mean wave is determined. The output of this is a signal that represents the correlation between the mean and all the waves. From here, we find the minimum correlation value across all time steps, which represents a human ID. From this human ID, we take the peak value and set this as the representative value for each user. When any test human ID is compared to the baseline ID, if the absolute value of the difference between both representative values surpasses a certain threshold (a certain percentage of the baseline ID's peak), then identification is positive, and the user is granted access. 
This threshold is a design parameter and is chosen to maximize the number of true positives (TP) while minimizing the number of false positives (FP). The full pipeline for this process can be seen in Fig. \ref{fig:human_ID_pipeline}.

\begin{figure*}[ht]
    \centering
    \includegraphics[width=0.8\textwidth]{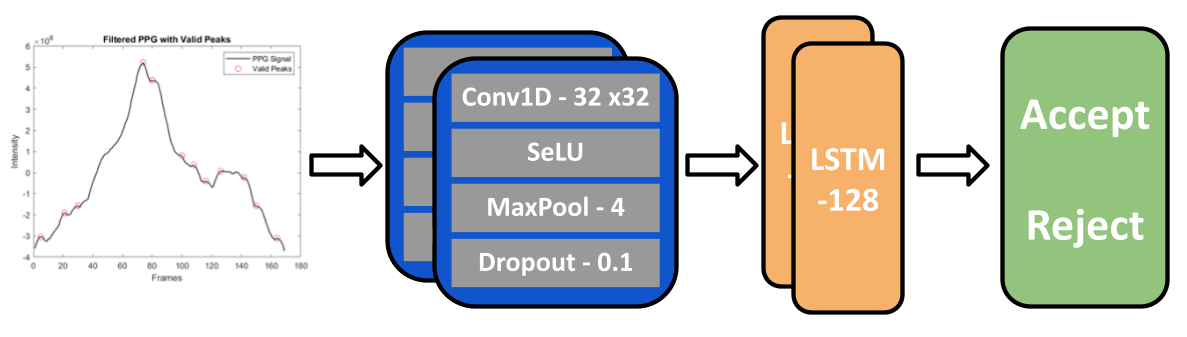}
    \caption{Deep-learning-based user identification pipeline from low frame-rate monochrome PPG signals. The system consists of two CNNs, two LSTMs, and a classification stage.}
    \label{fig:ML_ID_Pipeline}
\end{figure*} 

\subsubsection{Results}
\label{sssec:human_id_results}

We show the performance of the algorithm across different users in our dataset. For each user, we take their first trial and set this as the baseline. We then compare all other trials of the user (6 trials in total) to this baseline. From this, a perfect scenario occurs if the system correctly authenticates the user 6 times. To show the performance, we display the total number of true positives and false negatives, which should add up to 6 for each user. We also compare one trial of a user (e.g., user 1 trial 1) to all other trials of all the users except themselves (i.e., we compare user 1 trial 1 to all existing PPG signals except the PPG signals from user 1) to determine the number of false positives and true negatives. Ideally, there should be 24 true negatives.

Here, we present the results from the best and worst performing users. We start with user 2. 
We select trial 1 from user 2 and compare all of user 2's 6 trials to trial 1 of user 2, all of which are identified correctly. 
In addition, we compare trial 1 of user 2 to all other PPG signals in the database and we see that it correctly rejects (true negatives) all other 24 trials not belonging to user 2.

On the other hand, when this is repeated for user 3, we find that the system underperforms. 
While the system does a good job at rejecting impostor users (24 true negatives), it fails to correctly identify all the genuine attempts at authentication. Clearly, there is scope for further optimization.  

\subsection{Deep Learning-Based Method}
\label{ssec:deep_learning}

In this method, we utilize the deep learning approach, which uses a combination of CNNs and LSTMs \cite{Everson2018biometricnet}. 

\subsubsection{Algorithm}
\label{sssec:DL_algorithms}

In this approach, we start by filtering the PPG signal using a fifth-order moving average filter to remove noise. Then, we pass the data through two CNNs and two LSTMs, following which we send the output from these layers to a binary classification layer, which outputs a probability of the signal belonging to the user to be authenticated or not, as seen in Fig. \ref{fig:ML_ID_Pipeline}. The CNN layers perform one-dimensional convolution to extract features from the time series data, while the LSTM layers help extract time dependencies from the signal.

We use the following parameters for the two convolutional neural networks: 
\begin{itemize}
    \item Conv1D Layer:
    \begin{itemize}
        \item Number of filters: 32
        \item Filter size: 32 x 32
    \end{itemize}
    \item Activation Function: SeLU 
    \item MaxPool size: 4
    \item Dropout rate: 0.1 
\end{itemize}

For the LSTM deep network, we used 128 LSTM units and the \texttt{tanh} activation function. 
Furthermore, we use the following parameters throughout:  
\begin{itemize}
    \item Optimizer: Root Mean Square Propagation
    \item Loss Function: Categorical Crossentropy
    \item Number of Epochs: 100
    \item Batch size: 25
\end{itemize}

Finally, in the classification stage, the model performs binary classification and makes its best prediction on what class the PPG signal belongs to (user to be authenticated or not). The activation function for the classification stage was the softmax function.

\subsubsection{Results}
\label{sssec:DL_results}

Since there are only 6 PPG signals for the positive class and 24 for the impostor class, we need to ensure that our training and test sets have a representation of both the positive and negative classes. In addition, we recognize that there is a class imbalance of 6:24, which we account for by applying class weights that are inversely proportional to the class frequencies.

After performing 5-fold cross-validation, we compute the accuracy, F1 score, precision, and recall of the system over each fold as well as an average. 
We find that the model tends to always either predict class 1 or always predict class 0, even when the test set has a combination of both class 0 and class 1 (see Table \ref{tab:5_fold_results}), indicating an issue with our class balancing. One reason the system could be failing might be due to how the class weights are balanced. 

\begin{table}[t]
\centering
\caption{Results from 5-Fold cross-validation for the deep learning algorithm discussed in Section \ref{ssec:deep_learning}. User 3 is selected as the user to be authenticated. }
\label{tab:5_fold_results}
\begin{tabular}{|c|c|c|c|c|}
\hline
\textbf{Fold} & \textbf{Accuracy} & \textbf{F1 Score} & \textbf{Precision} & \textbf{Recall} \\ \hline
1-2 & 0.1667 & 0.2857 & 0.1667 & 1.0000 \\ \hline
3-4 & 0.8333 & 0.0000 & 0.0000 & 0.0000 \\ \hline
5   & 0.3333 & 0.5000 & 0.3333 & 1.0000 \\ \hline
\textbf{Mean} & \textbf{0.4667} & \textbf{0.2143} & \textbf{0.1333} & \textbf{0.6000} \\ \hline
\end{tabular}
\end{table}

\section{Conclusions and Future Work}
\label{sec:conclusions}
In this paper, we demonstrated that a low frame rate, monochrome, and off-the-shelf fingerprint sensor can effectively capture PPG signals for both vital sign estimation and user identification. We show that the heart rate can be accurately estimated with this fingerprint sensor and our proposed algorithm. Additionally, the user identification algorithm showed that the PPG signals from this fingerprint sensor could successfully serve as a secondary biometric authentication layer, adding liveness detection and reducing spoofing risks.

Future work will focus on expanding the dataset to include diverse user groups and further optimizing the algorithms. These advancements will enhance security by providing efficient biometric monitoring.

\bibliographystyle{IEEEtran}
\bibliography{src}

\end{document}